# Evidence for Magnetoreception in Red Drum (*Sciaenops ocellatus*), Black Drum (*Pogonias cromis*), and Sea Catfish (*Ariopsis felis*)

Joshua Courtney and Michael Courtney
BTG Research
9574 Simon Lebleu Road, Lake Charles, Louisiana, USA
Michael_Courtney@alum.mit.edu

**Abstract**
Over the past few decades, magnetoreception has been discovered in several species of teleost and elasmobranch fishes by employing varied experimental methods including conditioning experiments, observations of alignment with external fields, and experiments with magnetic deterrents. Biogenic magnetite has been confirmed to be an important receptor mechanism in some species, but there is ongoing debate regarding whether other mechanisms are at work. This paper presents evidence for magnetoreception in three additional species, red drum (*Sciaenops ocellatus*), black drum (*Pogonias cromis*), and sea catfish (*Ariopsis felis*), by employing experiments to test whether fish respond differently to bait on a magnetic hook than on a control. In red drum, the control hook outcaught the magnetic hook by 32-18 for $X^2 = 3.92$ and a P-value of 0.048. Black drum showed a significant attraction for the magnetic hook, which prevailed over the control hook by 11-3 for $X^2 = 4.57$ and a P-value of 0.033. Gafftopsail catfish (*Bagre marinus*) showed no preference with a 31-35 split between magnetic hook and control for $X^2 = 0.242$ and a P-value of 0.623. In a sample of 100 sea catfish in an analogous experiment using smaller hooks, the control hook was preferred 62-38 for $X^2 = 5.76$ and a P-value of < 0.001. Such a simple method for identifying magnetoreceptive species may quickly expand the number of known magnetoreceptive species and allow for easier access to magnetoreceptive species and thus facilitate testing of magnetoreceptive hypotheses.

## 1. Introduction

Magnetoreception is the sensory ability to detect a magnetic field and has been experimentally supported in certain species of bacteria, mollusks, insects, amphibians, birds, reptiles, mammals, and fish. (Johnsen & Lohmann, 2005) Evidence has accumulated steadily that some species of fish can derive positional information from the magnitude and inclination of the earth's magnetic field. (Kirschvink, 1989; Walker et al., 2003; Walker et al., 2006; Hart et al., 2012) To date, magnetoreception has only been demonstrated in a limited number of fish species.

It was once suggested that magnetite plays a key role in sensory perception of magnetic fields in all vertebrates that demonstrate magnetoreception (Diebel et al., 2000), but that conclusion has been called into question by elasmobranch experiments which seem to attribute magnetoreception to the ampullae of Lorenzini (O'Connell, 2011) and by experiments failing to find intracellular magnetite using sensitive detection techniques in a species known to be magnetoreceptive (Edelman, 2015). Identification of additional fish species demonstrating magnetoreception may be useful for future studies into sensory mechanisms and physiology.

The hypothesis that magnetoreception in elasmobranchs occurs via electromagnetic induction in the ampullae of Lorenzini gives rise to an expectation that teleosts cannot sense magnetic fields, even though several species of teleosts have been shown to be magnetoreceptive. (Courtney et al., 2014) The adoption of magnetic deterrents to reduce shark bycatch depends upon deterrent selectivity to only deter unwanted species and not reduce the catch of target species. Determining whether or not target teleost species are sensitive to magnets requires experiments in a range of teleost species. O'Connell and He have published results of testing for magnetoreception in a small number of teleost species (O'Connell and He, 2014). Though their results were negative, many more species remain to be studied, and their results should also be replicated by independent parties without commercial interests in magnetic hooks.

The present study tests whether the red drum (*Sciaenops ocellatus*), black drum (*Pogonias cromis*), gafftopsail catfish (*Bagre marinus*), and sea catfish (*Ariopsis felis*) have a preference for or against bait presented on a magnetic hook. These species were targeted, because they are relatively abundant and catching sufficient numbers was likely within a reasonable time in the study area. Fishing techniques were tailored for red drum, because this species is abundant in both nearshore Gulf and Atlantic waters off the United States where magnets may one day be used to reduce shark bycatch (O'Connell and He, 2014). Catch of this important species is prohibited in federal waters of the United States, so it was of interest to determine whether magnetic hooks would reduce or enhance catch rates.

## 2. Method and Materials

The method is patterned after the hook and line portion of O'Connell et al. (2011) with minor changes as necessary to ensure adequate numbers of specimens for statistical significance.  Magnetic and non-magnetic control hooks (with a lead sinker replacing the magnet) were fished simultaneously for equal times with the same bait and equivalent bait placement.  The resulting catch rates were then compared.  This method is not innovative, but allows for direct comparison with earlier studies in elasmobranchs, is inexpensive, and balances experimental accessibility and statistical significance in the likely catch.

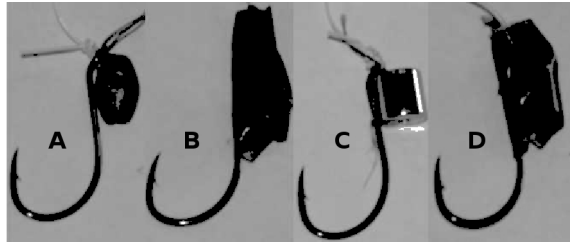

*Figure 1. Construction of magnetic hooks and lead control hooks used in the experiment on red drum, black drum and gafftopsail catfish.  A and B show the control hook design.  C and D show the magnetic hook design.  Hooks were fished with the magnet or lead weight covered with black tape (B and D).*

Magnetized hooks had a neodymium-iron-boron magnet attached.  Axially magnetized cylinder magnets approximately 12.5 mm long, 12.5 mm outer diameter, and 3 mm inner diameter (K&J Magnetics, Part # R828) were attached to 7/0 Gamakatsu Octopus J hooks (Part # 02417) as shown in Figure 1.  Magnets with a typical surface flux from 10,000 – 14,000 gauss were used.  These magnetic hooks produce a magnetic field comparable with the earth's field (0.5G) at a distance of 28 cm.  Therefore, fish possessing a magnetic sense capable of detecting the earth's field should be able to detect the magnetic field of the hook from a distance closer than 28 cm away.  Greater field sensitivity would extend the detection distance, but since magnetic dipole fields fall off as distance cubed, doubling the detection distance would require increased sensitivity by a factor of eight.  Control hooks were constructed by replacing the magnet with a 14 gram lead weight.  The top part of the hook with the magnet or lead control was then covered in black duct tape to reduce the visual differences between hooks.  Magnets and magnetic hooks were stored separately from non-magnetic hooks with care as not to impart any residual magnetism to the non-magnetic hooks.

In the red drum, black drum, and gafftopsail catfish experiment, two to four experienced anglers simultaneously fished two rods (one magnet and one control) with identical baits (crab or cut mullet) from a boat or beach location in the Gulf of Mexico near the Calcasieu Estuary in southwest Louisiana (USA), where the anglers have high success rates catching the target species.  Every time a fish was caught or a hook was rebaited, the positions of the two rods were reversed so that the magnetic hook and the control were fished in alternate locations in turn.  Every time a fish was caught, the angler called out the species and whether the catch occurred on the magnetic hook or the control, which were then recorded.

The hook size 7/0 was appropriate for the red drum, black drum, and gafftopsail catfish experiment, because specimens caught with these techniques and locations range from 60 cm to 100 cm in length. In contrast, sea catfish are much smaller, typically 15 cm to 40 cm in length. These fish are known to be abundant near certain piers in the Calcasieu Estuary and are readily caught on shrimp. For this species, a short shanked #1 size Mustad hook (part number 9174) was used by threading the 40 lb monofilament leader through the magnet or lead control and knotting the line just above it to prevent the magnet or control from sliding up the line away from the hook, as shown in Figure 2. Even though the hooks were smaller, the magnet lead weight used in the control hook were the same size as in the experiment in red drum, black drum, and gafftopsail catfish. Three experienced anglers fished two rods simultaneously and swapped the bait positions each time a fish was caught or a hook was rebaited. Sea catfish were so abundant that the hook positions were swapped every few minutes and the target of 100 sea catfish was caught in just a few hours.

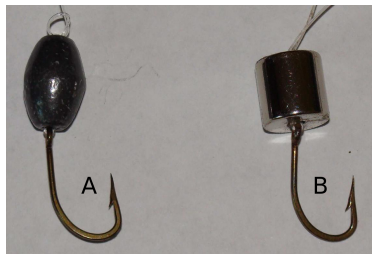

Figure 2. Control (A) and magnetic hooks (B) used in the sea catfish experiment.

## 3. Results

The number of red drum, black drum, and gafftopsail catfish caught on the magnet and control are shown in Figure 3 and Table 1. For red drum, 32 fish were caught on the control, and 18 fish were caught on the magnetic hook, yielding a Chi-square ($X^2$) value of 3.92 and a P-value of 0.048 relative to the expectation that equal numbers of fish would be caught on the magnet and the control. This suggests the difference in catch rates is statistically significant, supporting a finding of magnetoreception in red drum and that red drum tend to avoid magnetic hooks. All but one red drum caught in the study were "bull" red drum, meaning those over 68.6 cm (27 inches) in total length and likely to be sexually mature.

For black drum, 3 fish were caught on the control, and 11 fish were caught on the magnetic hook. In spite of the small total number, the strong ratio favoring magnetic hooks yields a $X^2$ value of 4.57 and a P-value of 0.033. This suggests the difference in catch rates is statistically significant, supporting a finding of magnetoreception in black drum with a preference toward magnetic hooks. All of the black drum caught in the study were "bull" black drum, meaning those over 68.6 cm (27 inches) in total length and likely to be sexually mature. The catch rate for black drum was small, because the black drum population in Calcasieu estuary has sharply declined since the overharvesting of oysters (a key food and habitat) in the area in 2010.

For gafftopsail catfish, 31 fish were caught on the control, and 35 fish were caught on the magnetic hook yielding a $X^2$ value of 0.24 and a P-value of 0.623. This suggests the difference in catch rates is not statistically significant and that gafftopsail catfish show no preference for either the control or the magnetic hook. All of the gafftopsail catfish caught in this study were between 40 cm and 80 cm in total length, suggesting they were most likely sexually mature.

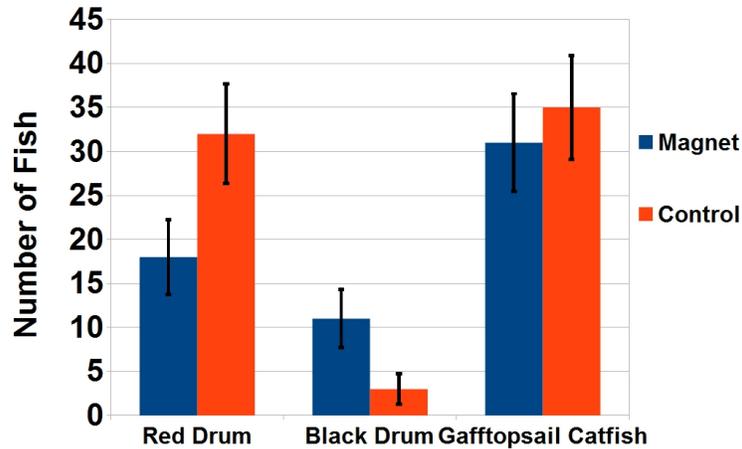

Figure 3. Number of red drum, black drum, and gafftopsail catfish caught on magnetic hook and control. Error bars show estimated uncertainty as the square root of number of observations in each case.

In the sea catfish experiment, 62 fish were caught on the control, and 38 fish were caught on the magnetic hook, for $X^2$ = 5.76 and a P value < 0.001. This suggests the difference in catch rates is statistically significant, supporting a finding of magnetoreception in sea catfish and that sea catfish tend to avoid magnetic hooks. The sea catfish caught in this experiment ranged in total length from 15 cm to 40 cm.

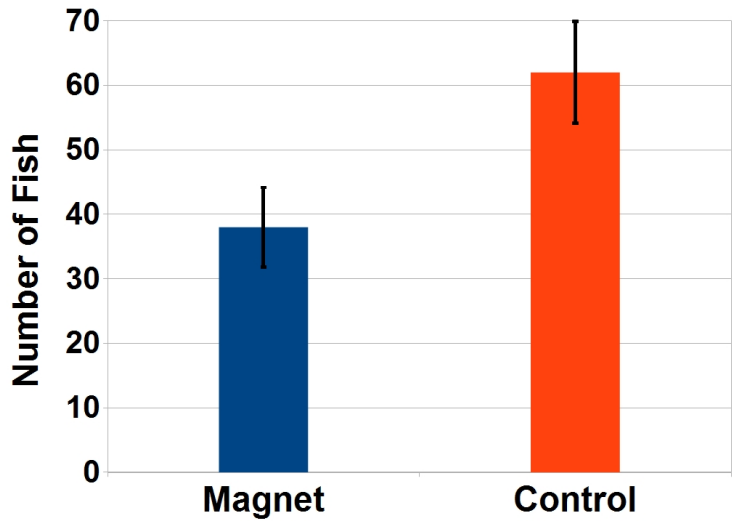

*Figure 4. Sea catfish caught on magnetic hook and control.*

*Table 1. Number of fish caught on each type of hook.*

| Species | Total Number ($n$) | Control | Magnet | $X^2$ | P |
|---|---|---|---|---|---|
| Red Drum | 50 | 32 | 18 | 3.92 | 0.048 |
| Black Drum | 14 | 3 | 11 | 4.57 | 0.033 |
| Gafftopsail Catfish | 66 | 31 | 35 | 0.242 | 0.623 |
| Sea Catfish | 100 | 62 | 38 | 5.76 | <0.001 |

## 4. Discussion

The results provide positive evidence for magnetoreception in red drum, black drum, and sea catfish, but do not show any preference for or against a magnetic hook in gafftopsail catfish. In addition to being the first indication of magnetoreception in three species, these results have practical applications both in fisheries management and with respect to future studies investigating the anatomy and physiology of vertebrate magnetoreception in more detail.

The absence of hook preference in gafftopsail catfish may indicate that gafftopsail catfish cannot detect the magnet, or it may mean that their feeding habits were not altered by detecting the magnetic hook. Our (unpublished) data over the past five years shows that gafftopsail catfish and other benthic feeding species in Calcasieu estuary are 10-15% below their expected weights and may be feeding more aggressively due to a lack of food.

It may be that black drum are the first species demonstrating increased catch rates for magnetic hook treatments.  (O'Connell et al., 2011; O'Connell and He 2014)  The strength of the magnetic hook suggests a fish capable of detecting the earth's geomagnetic field would also detect the hook from a distance of 28 cm.  For fish where the magnetic hook is a deterrent, every fish will be within 28 cm of the magnet before attempting to swallow the bait, thus fish capable of detecting the geomagnetic field will always detect the magnetic field of the hook before taking the bait.  In contrast, since the lower detection threshold of black drum is unknown, the effective detection radius of the magnetic hook treatment cannot be estimated.

We hypothesize that the magnetic field attracts black drum to the bait in cases where the fish happens to swim within the detection radius.  Black drum are known to be heavy feeders on mollusks.  They have powerful pharyngeal teeth and are capable of crushing prey including adult eastern oysters (*Crassostrea virginica*).  It may be that black drum have learned to associate the presence of a magnetic field with some food source.

Red drum are a popular game fish, but commercial fishing has been banned in federal waters of the United States and in the waters of many states as well.  Thus, use of magnetic hook treatments as shark deterrents may act as an effective red drum deterrent as well, keeping an important game fish from adding to the bycatch mortality of commercial long lines and trot lines.

In contrast, black drum are important commercial fish and are often targeted with inshore trot lines.  (Black drum are the top commercial fin fish in Louisiana, and their abundance and use in preparing artificial crab for retail markets supports an important commercial fishery in several other Gulf states.)  Sea catfish are viewed as a nuisance species since they are of little commercial value and prevent black drum from being caught once they occupy a hook on a commercial trot line.  Since the magnetic hook treatment seems to increase black drum catch rates and reduce sea catfish catch rates, there may be benefits to using magnetic hooks on inshore trot lines targeting black drum.  However, baits (crab chela), locations, and methods used in black drum trot lines produce negligible shark bycatch, so there is really no need for shark deterrents on black drum trot lines.  As a practical matter, the strength of the magnets used in this study tend to stick readily to each other, so some force is needed to separate magnetic hooks from each other.  In addition to the cost, this would add considerable risk and inconvenience to using them in large numbers in commercial trot lines and/or long lines where operators often set thousands of hooks each day.

Many of the black drum and red drum in this study ranged from 80-100 cm in total length and 10-15 kg in mass.  These are likely among the largest teleost fishes to be identified as magnetoreceptive.  The anatomy and physiology of magnetoreception are still something of a mystery.  Many sensory systems in fish (eyes, otoliths, lateral lines, ampullae of Lorenzini, etc.) scale with the size of the fish.  If the magnetoreceptive organs in these species also scale with fish size, they may be easier to locate and study.  Specimens are easily obtained.

Many factors may influence the effectiveness of magnetoreception in fish, including age, sex, length, weight, current, salinity, temperature, etc. Considering species as a whole when initially establishing magnetoreception is a standard experimental approach (O'Connell, 2011), as the purpose of this study was to establish the existence of magnetoreception rather than its sensitivity based on various factors, which may be the subject of future studies.

The most significant implication of this study may not be discovery of magnetoreception in three new species of fish, but rather the identification of magnetic hooks as an easy and convenient way of identifying magnetoreceptive species of fish for further study in better understanding the anatomical and physiological reception mechanisms.

## 5. Acknowledgments

We are grateful to BTG Research and the United States Air Force Academy for funding this work. We have incorporated valuable feedback from two peer reviewers. Elya Courtney and Elijah Courtney provided important input into the study design and assisted with the experiments.

## 6. References

Courtney, J., Courtney, Y. & Courtney, M. (2014). Review of Magnetic Shark Deterrents: Hypothetical Mechanisms and Evidence for Selectivity. *Aquatic Science and Technology*, 3(1), 70-81.   http://dx.doi.org/10.5296/ast.v3i1.6670

Diebel, C.E., Proksch, R., Green, C.R., Neilson, P., & Walker, M.M. (2000). Magnetite Defines a Vertebrate Magnetoreceptor, *Nature.* 406, 299-302.
 http://dx.doi.org/10.1038/35018561

Edelman, N. B., Fritz, T., Nimpf, S., Pichler, P., Lauwers, M., Hickman, R. W., ... & Keays, D. A. (2015). No evidence for intracellular magnetite in putative vertebrate magnetoreceptors identified by magnetic screening. *Proc. Natl. Acad. Sci.* U. S. A., 112(1), 262-267.   10.1073/pnas.1407915112

Hart, V., Kusta, T., Nemec, P., Blahova, V., Jezek, M., Novakova, P., Begall, S., Cerveny, J., Hanzal, V., Malkemper, E.P., Stipek, K., Vole, C., & Burda, H. (2012). Magnetic Alignment in Carps: Evidence from the Czech Christmas Fish Market. *PLOS ONE* 7(12), e51100.

Johnsen, S., Lohmann, K.J. (2005). The Physics and Neurobiology of Magnetoreception. *Nat. Rev. Neurosci.* 6, 703–712.
http://www.nature.com/nrn/journal/v6/n9/full/nrn1745.html

Kirschvink, J.L. (1989). Magnetite Biomineralization and Geomagnetic Sensitivity in Higher Animals: An Update and Recommendations for Future Study. *Bioelectromagnetics,* 10, 239-259. http://dx.doi.org/10.1002/bem.2250100304

O'Connell, C.P., Abel, D.C., Stroud, E.M., & Rice, P.H. (2011). Analysis of permanent magnets as elasmobranch bycatch reduction devices in hook-and-line and longline trials. *Fisheries Bulletin*. 109, 394-401.

O'Connell, C.P., & He, P. (2014). A Large Scale Field Analysis Examining the Effect of Magnetically-Treated Baits and Barriers on Teleost and Elasmobranch Behavior. *Ocean & Coastal Management*. 97, 29-37.

Walker, M.M., Diebel, C.E., & Kirschvink, J.L. (2003). Detection and Use of the Earth's Magnetic Field by Aquatic Vertebrates. In S.P. Collin & N.J. Marshall (Eds.), *Sensory Processing in Aquatic Environments* (53–74). New York, NY: Springer-Verlag.

Walker, M.M., Diebel, C.E., & Kirschvink, J.L. (2006). Chapter 8: Magnetoreception. In: T. Hara & B. Zielinski (Eds.), *Sensory Systems Neuroscience: Fish Physiology.* 25, 335-374.